\documentclass[a4paper,12pt]{article}
\pdfoutput=1
\usepackage{amsmath}
\usepackage{graphicx}
\usepackage{amssymb}
 \DeclareMathOperator{\Tr}{Tr}
 
\DeclareMathOperator{\Pexp}{Pexp}

\title{Nonlocal field correlators on the lattice in HP$^1$ $\sigma$-model}

\author{V.\,D.\,Orlovsky\/\thanks{e-mail: orlovskii@itep.ru},
V.\,I.\,Shevchenko}

\date{ Institute for Theoretical and Experimental Physics, \\ B.
Cheremushkinskaya 25, 117218 Moscow, Russia}

\begin{document}

\maketitle

\abstract{Connected two-point field strength correlators have been
measured on the lattice in quaternionic projective $\sigma$-model
of pure SU(2) Yang-Mills theory. The correlation lengths,
extracted from the exponential fit for these correlators, are
found to be $\lambda_1^{-1} = 1.40(3)$\,GeV and $\lambda^{-1} =
1.51(3)$\,GeV in good agreement with other existing calculations.
The dependence of bilocal functions on the connector shape was
studied.}

\newpage

The studies of nonperturbative (NP) aspects of QCD are well known
to be of prime importance. This is the most interesting and rich
field, albeit the most complicated one. It is commonly believed
that the origin of complex NP phenomena such as confinement,
chiral symmetry breaking etc is highly nontrivial structure of QCD
vacuum. The important step in the NP vacuum fields investigations
was undertaken in the seminal paper \cite{SVZ} where the gluon
condensate -- NP average of gluon fields over vacuum state -- was
introduced. This object, still playing essential role in studies
of QCD vacuum structure is defined as
\begin{equation}\label{G_2}
G_2 = \frac{\alpha_s}{\pi}\left\langle F^a_{\mu\nu}(0)
F^a_{\mu\nu} (0) \right\rangle,
\end{equation}
where $F^a_{\mu\nu}$'s are the field strength tensors. In the QCD
sum rules approach this is universal quantity (together with
chiral condensate $\langle \bar{\psi} \psi\rangle$ and a few
higher NP averages), describing QCD vacuum. The sum rule method
and idea about NP condensates have proven to be very fruitful.
Using this technique masses of resonances, decay constants and
other quantities of phenomenological interest were computed in
good agreement with experiment. However, there are questions,
which are difficult to answer in the sum rule method. In general,
they arise when one is interested in such objects as potentials,
excited states etc, i.e. for physical situations when one of the
space-time scales characterizing the problem in question is large.
It was found (in \cite{DiG_Panag} on the lattice and later by
other authors), that the word "large" here has rather precise
meaning: there is an important parameter, characterizing NP
dynamics of vacuum fields -- correlation length, which defines the
spacial decay of connected gauge-invariant bilocal correlators of
field tensors. It is natural to expect that original sum rule
method, operating with local quantities, is not applicable to the
situations where nonlocal properties of vacuum fields are
essential.

The method of field correlators (MFC) \cite{DoschSimonov, Dosch,
Simonov} (see \cite{DiGDSS} for review) can be considered as a
development of sum rules method explicitly taking into account
long distance effects. The dynamical input is a set of NP field
strength correlators
\begin{equation}\label{FC}
\Delta_{\mu_1\nu_1, \ldots, \mu_n\nu_n} = \Tr \left \langle
gF_{\mu_1\nu_1}(x_1) \Phi (x_1,x_2) gF_{\mu_2\nu_2}(x_2) \ldots
gF_{\mu_n\nu_n}(x_n) \Phi (x_n,x_1) \right \rangle,
\end{equation}
where $\Phi(x,y) = \Pexp \left( \int \limits_y^x A_\mu dz_\mu
\right)$ with integration along some path, connecting the points
$x$ and $y$, are the phase factors, introduced for the sake of
gauge invariance. As a matter of principle these objects can be
understood as infinite power series of NP condensates and in this
sense their theoretical status is absolutely equivalent to
(\ref{G_2}). But it is more convenient as is seen below, to
express (\ref{FC}) in terms of correlation lengths.
Gauge-invariant observables for QCD processes of phenomenological
interest can be expressed through (\ref{FC}) via cluster
expansion. MFC includes not only perturbative QCD at small
distances, but also large distance nonperturbative effects (see
\cite{DiGDSS} and references therein).

It is remarkable, that one can describe most NP QCD phenomena with
high accuracy by means of the lowest two point correlator
$\Delta^{(2)}_{\mu_1\nu_1, \mu_2\nu_2}$, while higher cumulants
can be considered as small corrections \cite{DiGDSS}. In
particular, the direct consequence of this fact is the so called
Casimir scaling law for static potential between two quarks
\cite{Simonov_Casimir, SS_Casimir}, clearly seen on the lattice
\cite{Bali_Casimir}. The corresponding formalism is
called the Gaussian dominance approximation. Here we have single
fundamental input -- bilocal correlator, which has the following
parametrization \cite{DoschSimonov}:
\begin{multline}\label{Delta(2)}
\Delta^{(2)}_{\mu_1\nu_1, \mu_2\nu_2} = g^2\Tr \left \langle
F_{\mu_1\nu_1}(x) \Phi (x,0) F_{\mu_2\nu_2}(0)
 \Phi (0,x) \right \rangle = \\ = (\delta_{\mu_1\mu_2} \delta_{\nu_1\nu_2} -
 \delta_{\mu_1\nu_2}
 \delta_{\mu_2\nu_1})(D(x^2)+D_1(x^2))+ \\ +
 (x_{\mu_1}x_{\mu_2}\delta_{\nu_1\nu_2}
  - x_{\mu_1}x_{\nu_2}\delta_{\mu_2\nu_1}+ x_{\nu_1} x_{\nu_2}
  \delta_{\mu_1\mu_2}  -
 x_{\nu_1}x_{\mu_2}\delta_{\mu_1\nu_2})\frac{d}{dx^2}D_1(x^2),
\end{multline}
where $D(x^2)$ and $D_1(x^2)$ are some functions of distance
between two points and integration in $\Phi(x,0)$ goes along the
straight lines.

At short distances the expression (\ref{Delta(2)}) can be expanded
in powers of $x^2$, the corresponding coefficients happen to be
proportional the condensates of higher powers (see
\cite{Nik_Radyush, Mikhailov, Grozin, Dorokhov, SS_relations} for
details). The leading nonabelian term is given by (notice that our
normalization is different from that of the cited papers):
\begin{equation} \left.\frac{dD(x)}{dx^2}\right|_{x=0} = \frac{g^3}{96}
f^{abc}\langle F^a_{\mu\nu}F^b_{\nu\rho}F^c_{\rho\mu}\rangle
\label{pol}
\end{equation}
(thus at small distances the correlator is Gaussian as a function
of $x$). On the other hand the NP long distance contribution can
be parameterized by exponentials with the correlation lengthes
$\lambda$ and $\lambda_1$:
\begin{equation}\label{D and D1 as exp}
D(x) = A\exp(-|x|/\lambda), \quad D_1(x) =
A_1\exp(-|x|/\lambda_1).
\end{equation}
Our lattice results demonstrate that exponential behavior is a good
approximation starting just from the distance of order of the
correlation length $\lambda, \lambda_1$ (see details below), while
for smaller distances the regime (\ref{pol}) is expected (but not
explicitly seen by us due to insufficient lattice precision).

Very important source of information on field correlators are
lattice measurements \cite{DiG_Panag, Bali_Br}. There is an
immediate problem, however. Measuring (\ref{Delta(2)}) on the
lattice one would get $\sim 1/x^4$ piece at small distances as
predicted by perturbation theory together with some NP contributions like
(\ref{D and D1 as exp}) and extracting the NP part is a nontrivial
task. This problem can be solved by using cooling technique (as in
\cite{DiG_Panag}), which allows to eliminate short range
fluctuations, or by smearing procedure as in \cite{Bali_Br}.
Another possibility is to expand nonlocal correlators in powers of
local condensates, but needless to say that practically it is not
possible to study all the series of local quantities. So, it is
desirable to find some alternative approach, which allows to
single out the NP signal and to compare with the other lattice
methods.

For this purpose we can use an opportunity to consider modified
field configurations, close in some sense to the initial
configurations. A nice example of such modification is replacement
of the initial SU(2) gauge theory by configurations of scalar
fields representing nonlinear $\sigma$-model with target space
being quaternionic projective space HP$^1$. This was realized on
the lattice in \cite{Gubarev_Mor1,Gubarev_Mor2}, where both local
objects like gluon condensate and nonlocal quantities were
measured in good agreement with known results. So, the string
tension calculated in terms of HP$^1$ projected fields turned out
to be very close to full SU(2) string tension,
$[\sigma^{HP^1}/\sigma^{SU(2)}]^{1/2} = 1.04(3)$ in the continuum
limit. It was established, that this projection captures only NP
content of gauge background, cutting contribution from
perturbation theory. So, it is very natural to look at correlators
of gluon fields in this approach and this is done for the first
time in the present paper.

Due to lack of space we can not discuss all details of
constructing HP-projected fields (see \cite{Gubarev_Mor1} for this
purpose), but to be self-contained we give here the main steps.
The simplest explicit parametrization of HP$^1$ is provided by
normalized quaternionic vectors
\begin{equation}
|q \rangle = [q_0,q_1]^T, \quad q_i \in H, \quad \langle q|q
\rangle = \bar{q}_iq_i = 1 \in H,
\end{equation}
where H is the field of real quaternions. The states $|q \rangle$
describe 7-dimensional sphere, while the HP$^1$ space is the set
of equivivalence classes of $|q\rangle$ with respect to the right
multiplication by unit quaternions (elements of SU(2) group)
\begin{equation}
|q \rangle \sim |q \rangle v, \quad |v|^2=1, \quad v \in H.
\end{equation}
Configuration $|q_x \rangle$, assigned to the lattice as the best
possible HP$^1$ fields approximation provides the minimum of
functional
\begin{equation}
F(A,q) = \int(A_{\mu} + \langle q | \partial_\mu | q\rangle)^2
\end{equation}
for given SU(2) field $A_{\mu}$. The working tool for computations
is link variable
\begin{equation}
U_{x,\mu} = \frac{\langle q_x| q_{x+\mu} \rangle}{|\langle q_x |
q_{x+\mu} \rangle |}.
\end{equation}

What is actually measured is a loop made of two plaquettes
(smallest rectangular contours made of four link variables)
$P_{\mu\nu}(x) = U_{x,\mu}U_{x+\mu,\nu}U^+_{x+\nu,\mu}U^+_{x,\nu}$
connected by the phase factors (product of links $U_i$ between two
points) along straight lines, with different orientation of the
plaquettes:
\begin{equation} \label{corr_lattice}
\frac{1}{a^4} \left\langle \Tr(P_{\mu_1\nu_1}(x)-1) \left(
\prod\limits_i U_i \right) (P_{\mu_2\nu_2}(0)-1) \left(
\prod\limits_i U^+_i \right) \right\rangle_{HP^1},
\end{equation}
where $a$ is the lattice spacing.

There are two possibilities to obtain nonzero answer in
(\ref{Delta(2)}) -- when parallel planes $(\mu_1 \nu_1)$ and
$(\mu_2 \nu_2)$ are perpendicular to the vector $x$ ($D_\bot(x)$)
and when planes $(\mu_1 \nu_1)$ and $(\mu_2 \nu_2)$ are parallel
and $x$ lies in this plane ($D_\|(x)$):

\begin{gather}\label{D_tr and D_par}
D_\bot(x) = D(x)+D_1(x), \\ D_\|(x) = D(x)+D_1(x) + x^2
\frac{\partial D_1(x)}{\partial x^2}.
\end{gather}

One attractive feature this approach is its relatively low
computational cost. Lattice simulations have been performed on PC
1.66GHz on 100 configurations (taken from \cite{Fedor}) at $\beta
= 2.6$ corresponding to the lattice spacing $a=0.06$ fm. The total
volume was $40^4$. Other configurations with larger lattice
spacing were also used for scaling properties study.
Parametrization (\ref{D and D1 as exp}), which works very well
starting from two lattice spacing (remember that the HP$^1$
projected fields do not contain perturbation theory contributions
at small distances) gives for correlator $D_1(x)$ from (\ref{D_tr
and D_par}) (see Fig.1) rather stable result $\lambda^{-1}_1 =
1.40(3)$\,GeV (or $\lambda_1 = 0.14(1)$\,Fm) in good agreement
with \cite{DiGiacSU(2)}. The data for the correlator $D(x)$
(Fig.2) allow us to extract $\lambda^{-1}=1.51(3)$\,GeV (or
$\lambda = 0.13(1)$\,Fm). The preexponential factors $A \approx
0,11$\,GeV$^4$, $A_1 \approx 0,06$\,GeV$^4$. We studied the
scaling properties of correlation length and found that the
dependence on the lattice spacing is almost inessential for this
quantity (see Fig.3). On the other hand, the quantity $A+A_1$,
which is proportional to the condensate $G_2$, scales as
$\alpha/a^2+\beta$ (see Fig.4) with $\alpha \approx (95$\,MeV$)^2$
and $G_2=(6N_c/\pi^2)\beta = 0,062(6)$\,GeV$^4$ in complete
agreement with results of \cite{Gubarev_Mor2}. The quoted
uncertainties are the standard statistical errors. For reader's
convenience we give the values of gluonic correlation length
$\lambda$, obtained in other papers, in Table~1.
\\

Table 1. Correlation length $\lambda$.

\begin{tabular}{ccccc}
  \hline
   & \cite{DiGiacSU(2)}$_{SU(2)}$ & \cite{DiG_Panag}$_{SU(3)}$ & \cite{Bali_Br}$_{SU(3)}$ &  \cite{Simonov_Badalyan_Nefediev}$_{SU(3)}$ \\
  \hline
  $\lambda$, fm & 0.13 & 0.22 & 0.12 & 0.10 \\
  \hline
\end{tabular}
\\
\\

It is interesting to check how the correlators depend on the shape
of curve connecting the points $0$ and $x$. It turned out, that
correlator $D_\bot(x)$ holds the form close to exponential decay,
if we replace straight connector by the U-like curve (see Fig.5).
Fig.6 shows how that exponential index $\lambda$ depends on the
depth $l$ of the curve flexure. Unlike to this case of sizeable
changing of correlation length for the function $D_\bot(x)$ (by
$\sim 16\%$) correlator $D_\|(x)$ does not preserve exponential
form under the changing of the shape. We have also checked
how shape of the connector influences the integral quantities.
To this end, we consider the sum of two-point ``U-like" and
``L-like" correlators, taken over rectangle $R\times T$ in the
$(xt)$-plane
\begin{equation}\label{sum_rectang}
w_{U,L} = \sum\limits_{l,l'}\frac{1}{a^4} \left\langle
\Tr(P_{tx}(l)-1) \Phi_{U,L}(l,l') (P_{tx}(l')-1) \Phi_{U,L}(l',l)
\right\rangle_{HP^1}.
\end{equation}
The typical contributions to the sum are shown on Fig.7. The value
$w_{U,L}$ scales as the rectangle square $w_{U,L} \propto
\bar{\sigma}_{U,L}RT$ for sufficiently large rectangles. The
perimeter terms, also presented in $w_{U,L}$, may be taken into
account as $R$- and $T$-dependence of $\bar{\sigma}_{U,L}$. The
dependence of correlators on the connector shape shows as
difference between $\bar{\sigma}_{U}$ and $\bar{\sigma}_{L}$. The
sum (\ref{sum_rectang}) as function of rectangle square is shown
on Fig.8 at fixed $R=7$. As turned out the difference is small:
\begin{equation}
\frac{\bar{\sigma}_{L}-\bar{\sigma}_{U}}{\bar{\sigma}_{L}+\bar{\sigma}_{U}}
\approx 5\%,
\end{equation}
so the integral quantities are almost insensitive to changing of
the connector form. Another interesting question is the relation
of the coefficient $\bar{\sigma}_{U,L}$ to the string tension
$\sigma^{HP_1}$ (one would expect approximate equality of these
quantities in Gaussian stochastic model). It will be discussed
elsewhere \cite{preparation}.

Finally, it is natural to expect that correlators of pure electric and
pure magnetic fields
\begin{equation}\label{E_and_B}
\mathcal{E}(x) = \Delta^{(2)}_{i4i4}(x), \quad \mathcal{B}(x) =
\frac{1}{4}\epsilon_{ilm} \epsilon_{ijk} \Delta^{(2)}_{lmjk}(x)
\end{equation}
should coincide at $x=0$ because of $O(4)$
rotational symmetry of the problem. As we can see from Fig.9, it is indeed so --
two lines, corresponding to logarithms of electric and magnetic
correlators (\ref{E_and_B}) come together at small $x$ (the error
bars are within the symbols).

In conclusion, we have measured the two-points correlators of
vacuum fields on the lattice in HP$^1$ $\sigma$-model. This
approach has allowed us to get rid of the perturbation theory
contribution and to use the long-distance nonperturbative
expression in the whole region of distances, beginning from two
lattice spacing. Our measurements provide solid confirmation of
exponential behavior for long distances of functions $D(x)$ and
$D_1(x)$, parameterizing two-point correlators.  Small distance
behavior is the question of separate interest, but due to
insufficient precision of our lattice calculations we can not
catch expected Gaussian $x^2$-dependence and we do not concentrate
on this problem, being interested in long-distance characteristics
of two-point correlators. So, the corresponding correlation
lengthes are found to be $\lambda_1^{-1} = 1.40(3)$\,GeV and
$\lambda^{-1} = 1.51(3)$\,GeV in good agreement with other known
calculations using completely different lattice technique. It is
worth noticing that parameters $\xi = (A \lambda^4)^{1/2} \approx
0.14, \, \xi_1 = (A_1 \lambda_1^4)^{1/2} \approx 0.12$ turn out to
be rather small, confirming the MFC picture of NP vacuum fields
being weak in units of the inverse correlation length (see related
discussions in \cite{SS_confinement}). We also studied the
dependence of bilocal functions on the connector shape and found
some decrease of the correlation length with the increasing of the
connector length, corresponding to the function $D_\perp (x)$. In
the case of the function $D_{\|}(x)$ this dependence can be
established via the integral characteristic as the small
difference (about 5\%) between slopes of $w_U(S)$ and $w_L(S)$
(\ref{sum_rectang}) as functions of rectangle square.

The authors are grateful to F.\,V. Gubarev, M.\,I. Polikarpov and
Yu.\,A. Simonov for helpful discussions. The research was
supported by the grant for support of scientific schools
NS-4961.2008.2.

 \begin{figure}[p]
  \begin{center}
   \includegraphics[height=8cm]{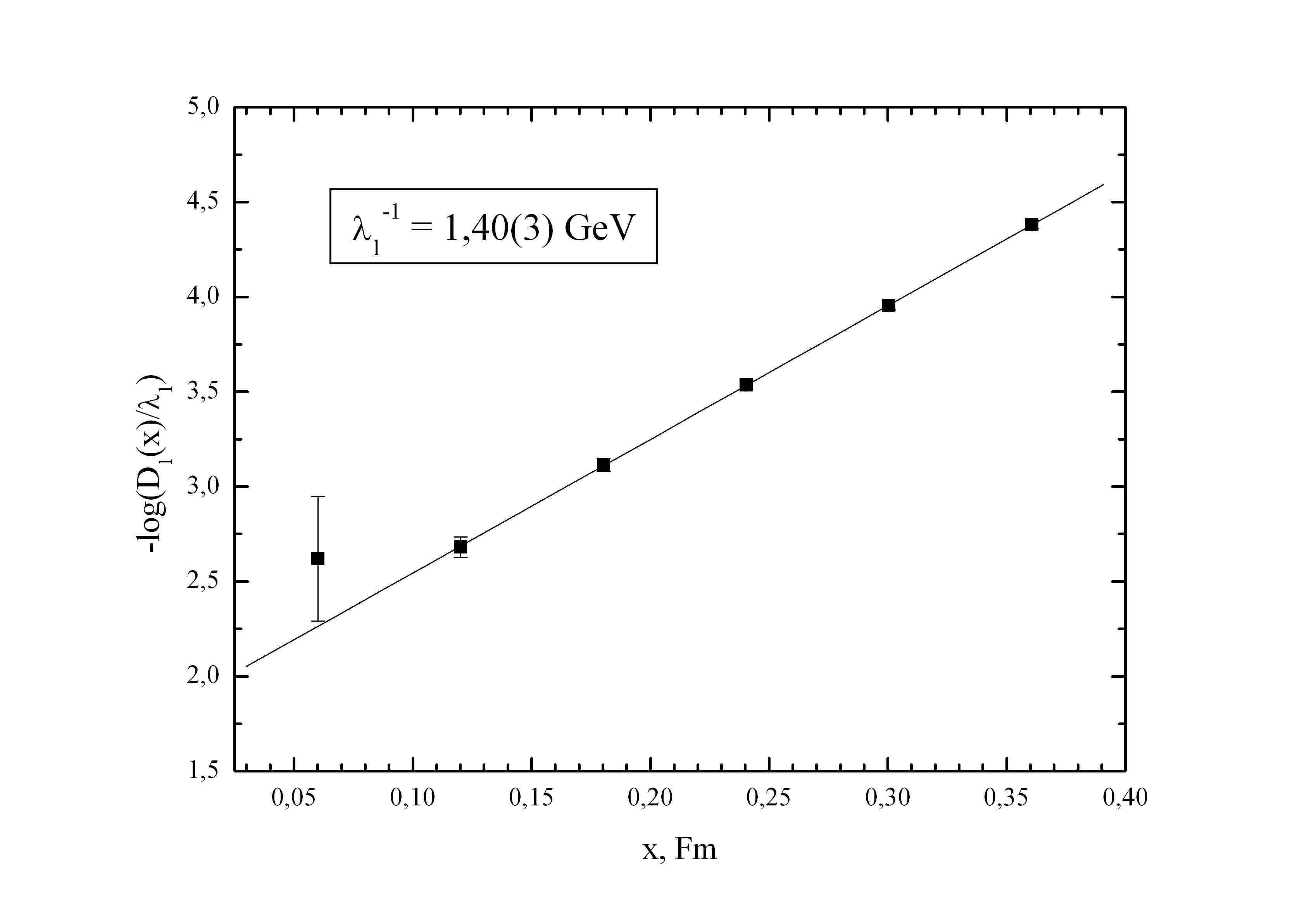}
  \end{center}
  \caption{Function $\log D_1(x)$ (conventional units) from the measurements of two point correlators (9)}\label{D1}
 \end{figure}

 \begin{figure}[p]
  \begin{center}
   \includegraphics[height=8cm]{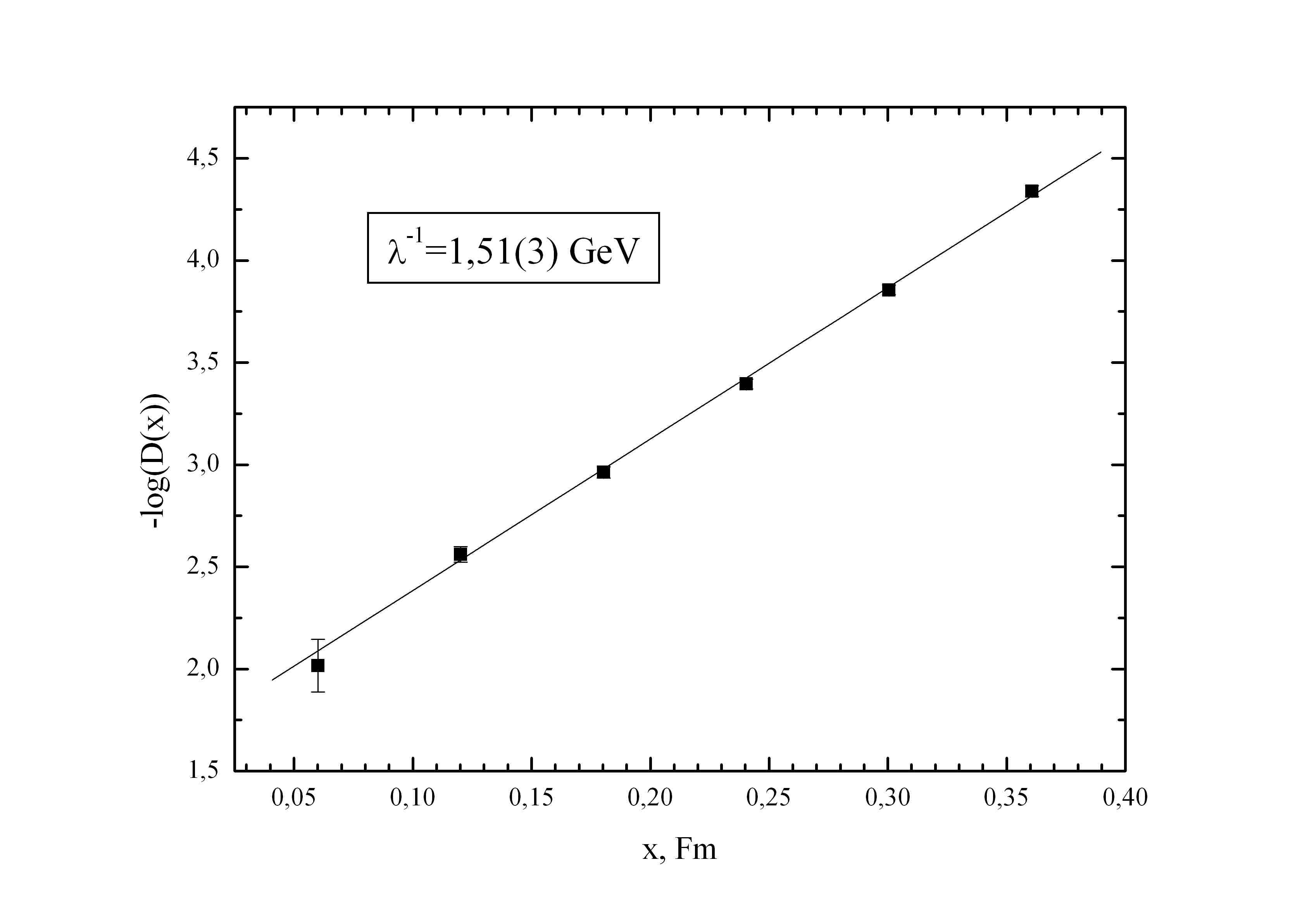}
  \end{center}
  \caption{Function $\log D(x)$ (conventional units) from the measurements of two point correlators (9)}\label{D}
 \end{figure}

 \begin{figure}[p]
  \begin{center}
   \includegraphics[height=8cm]{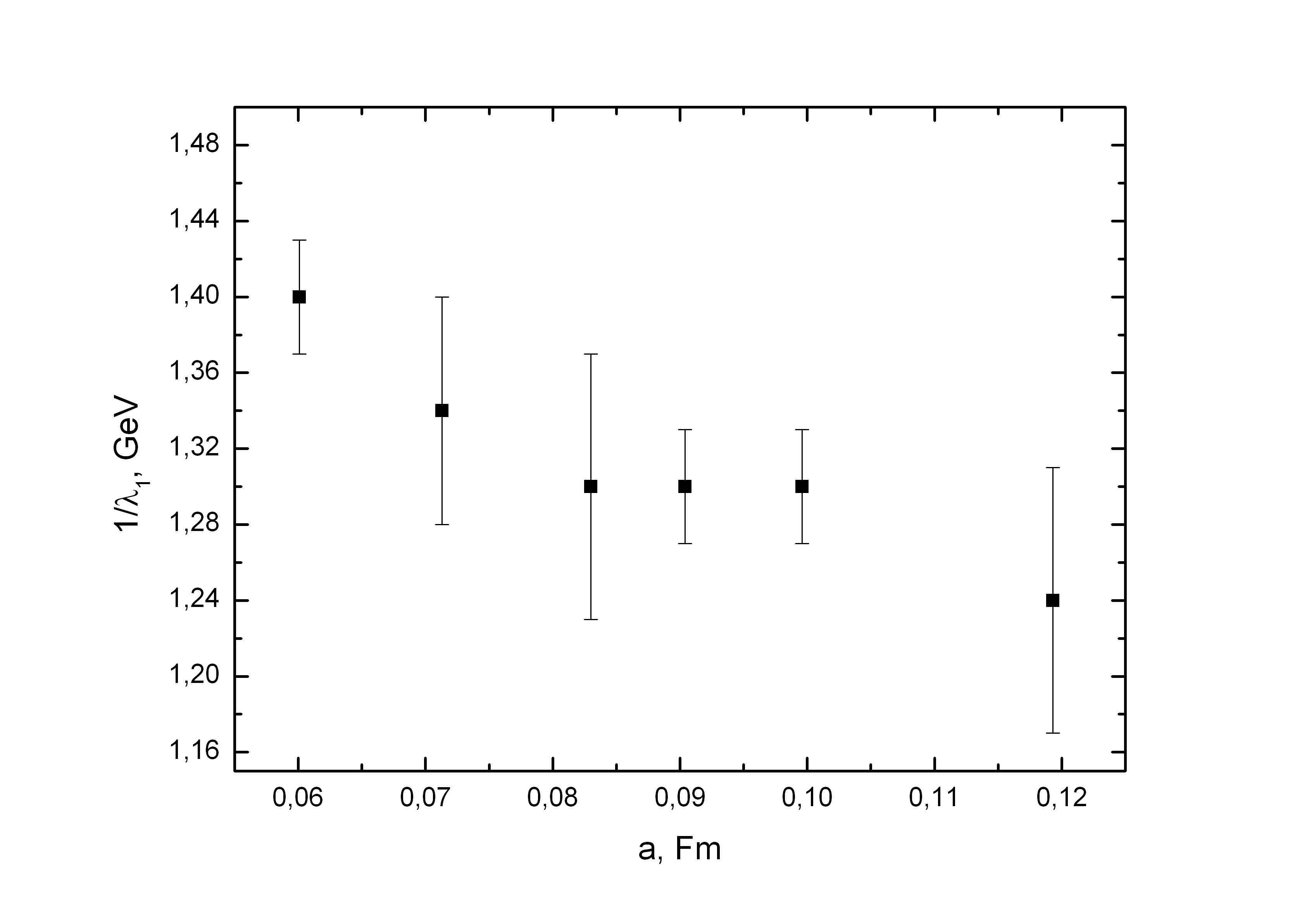}
  \end{center}
  \caption{Correlation length $\lambda_1$ as a function of lattice
spacing}\label{scaling}
 \end{figure}

 \begin{figure}[p]
  \begin{center}
   \includegraphics[height=8cm]{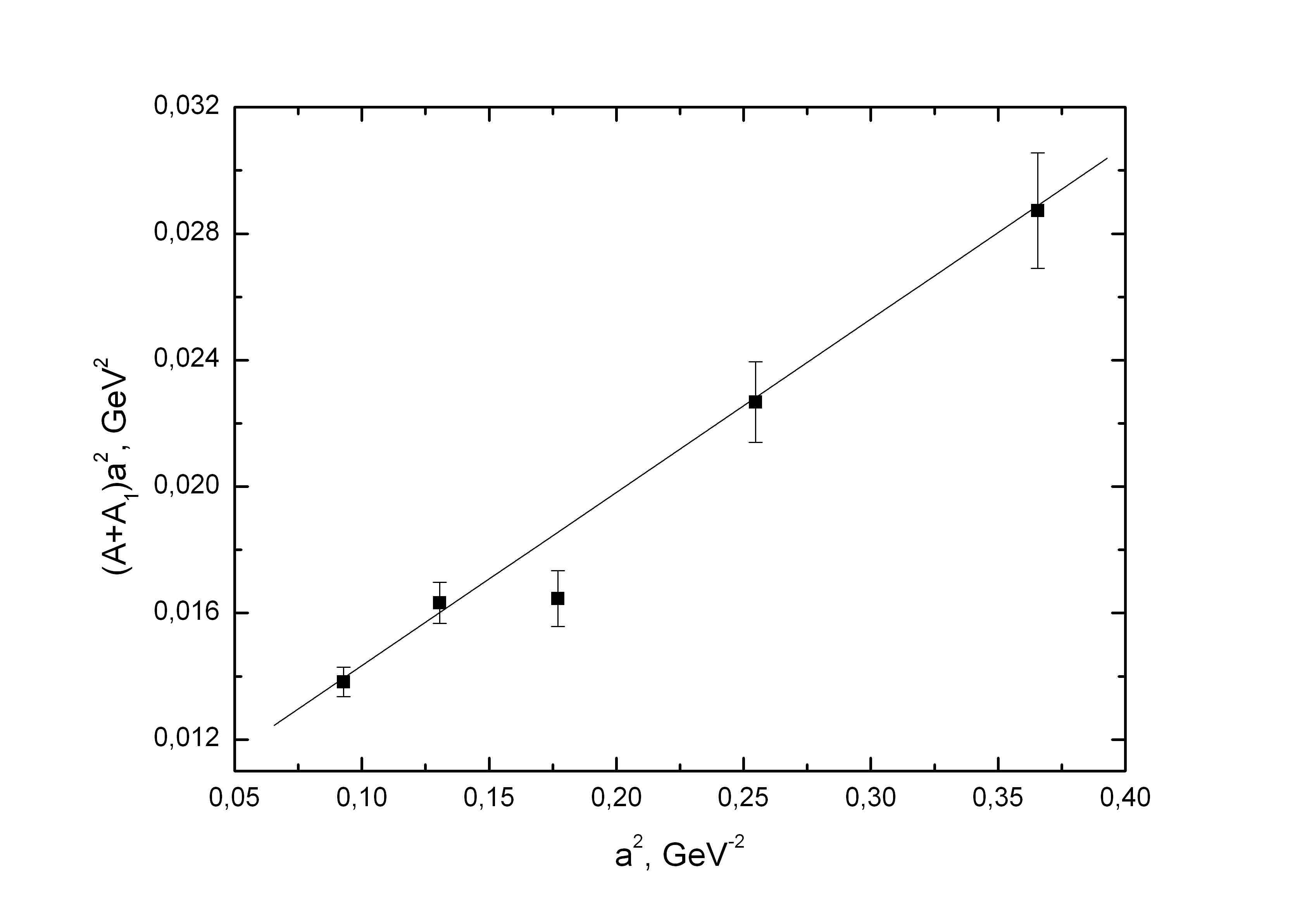}
  \end{center}
  \caption{Condensate $D(0)+D_1(0)$ as a function of lattice
spacing}\label{scaling(A+A1)}
 \end{figure}

 \begin{figure}[p]
  \begin{center}
   \includegraphics[height=3cm]{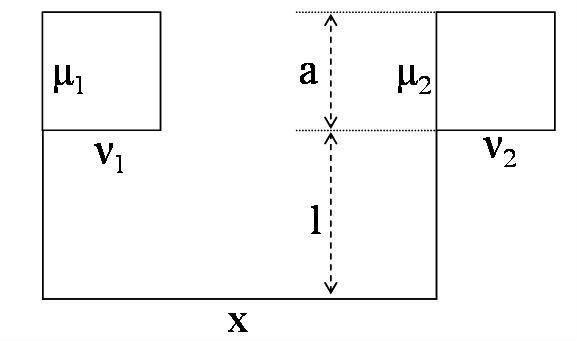}
  \end{center}
  \caption{Correlator on the lattice for U-like connector}\label{U-like}
 \end{figure}

 \begin{figure}[p]
  \begin{center}
   \includegraphics[height=8cm]{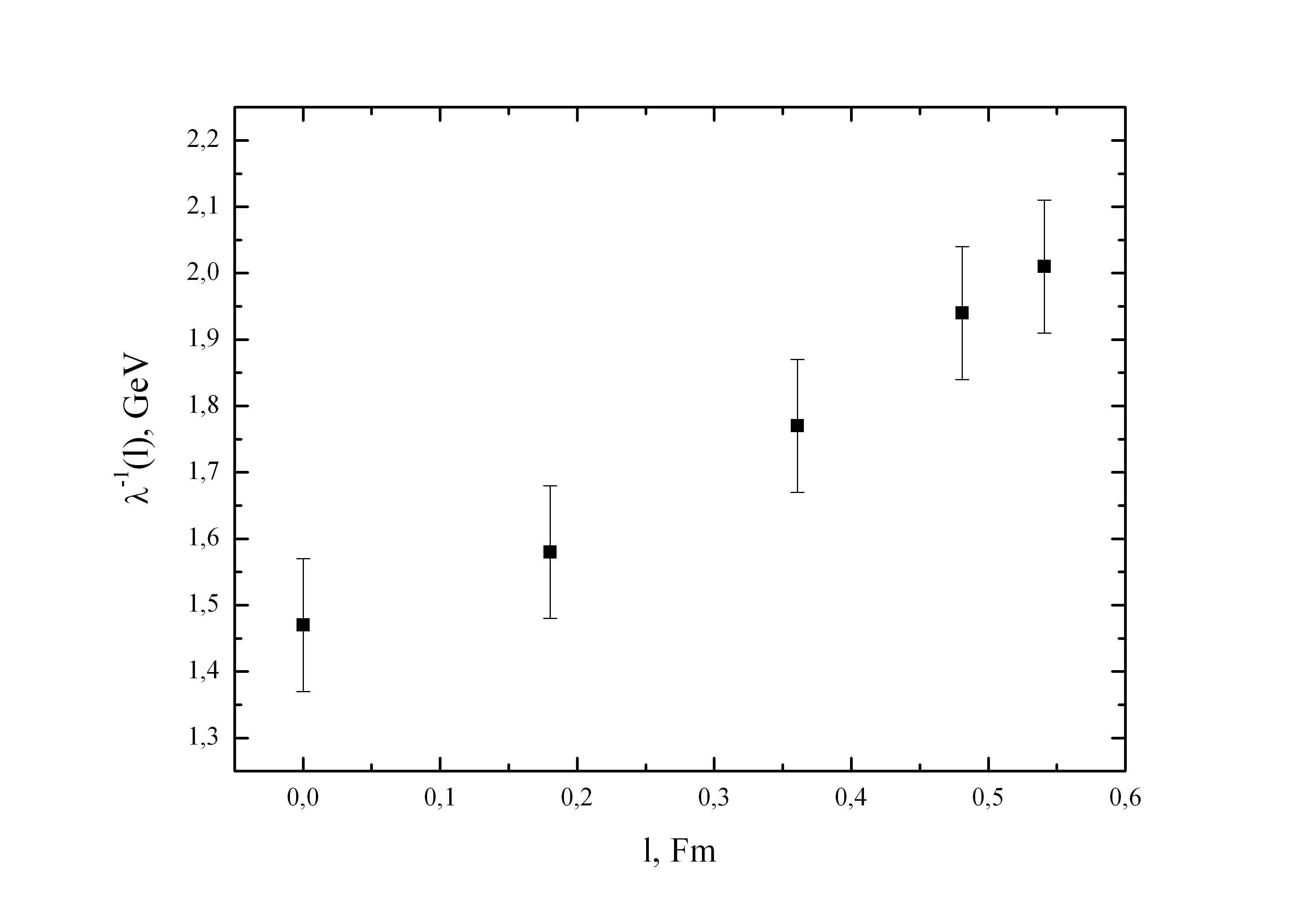}
  \end{center}
  \caption{Exponential index as a function of the connector shape}\label{lambda(l)}
 \end{figure}

 \begin{figure}[p]
  \begin{center}
   \includegraphics[height=4cm]{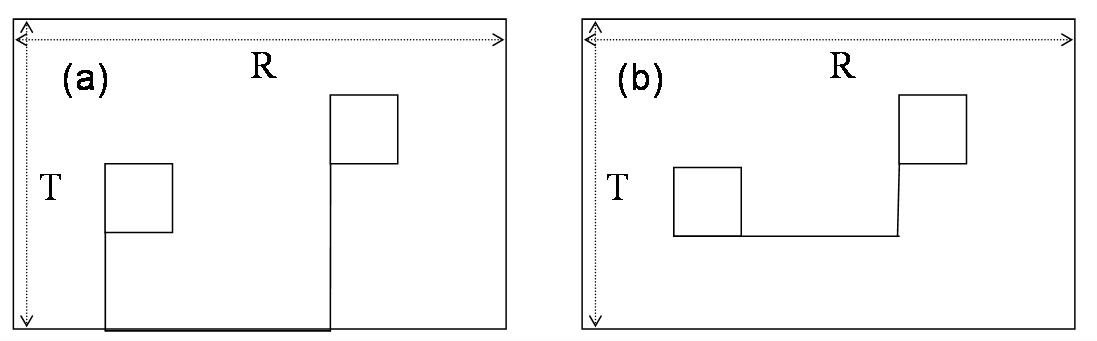}
  \end{center}
  \caption{Typical contribution to the sum (12) for (a)
``U"-like and (b) ``L"-like correlators}\label{sigma_graph}
 \end{figure}

  \begin{figure}[p]
  \begin{center}
   \includegraphics[height=8cm]{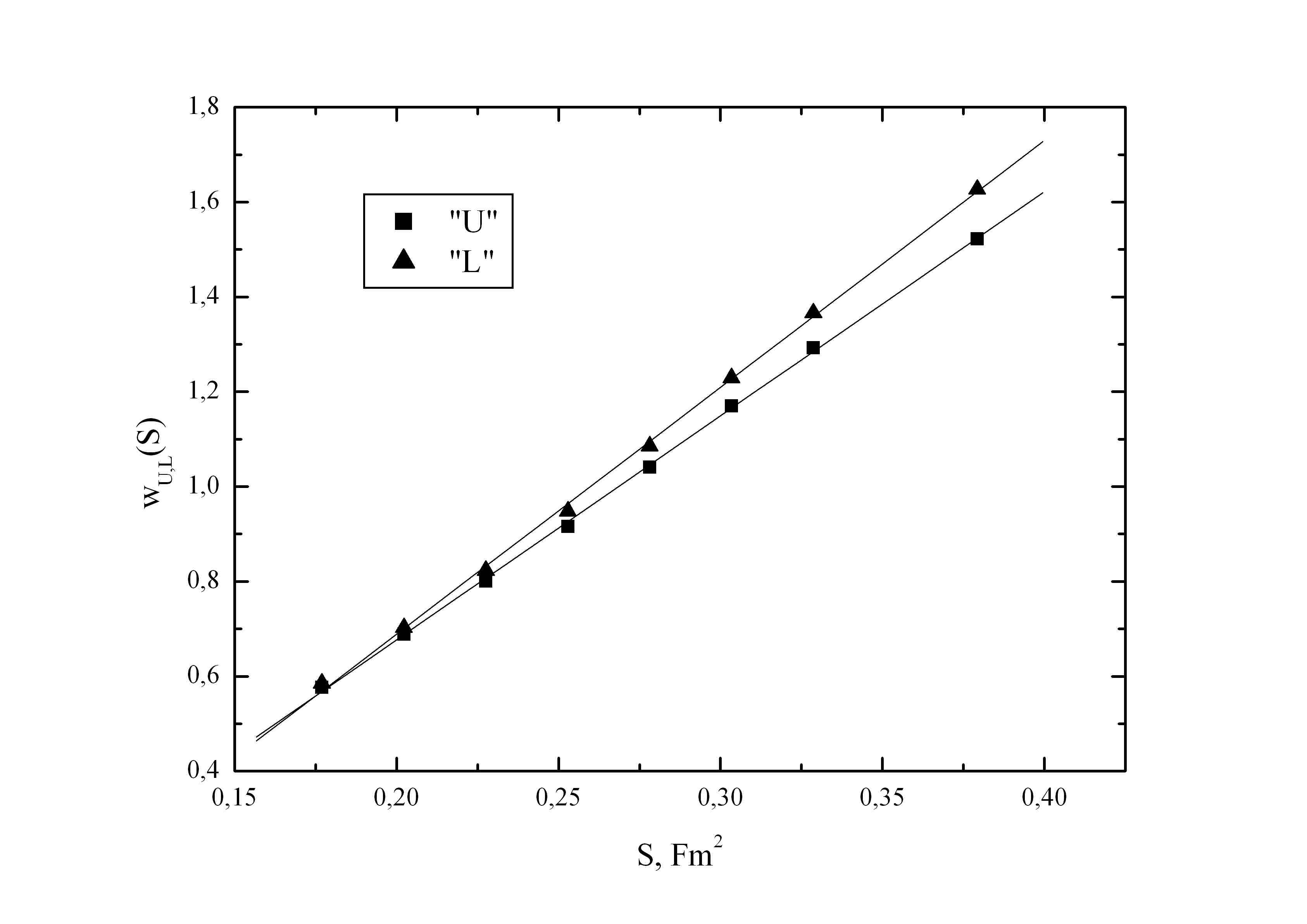}
  \end{center}
  \caption{The sum (12) as function of rectangle square $S=(R=7)\times T$ for ``U"- and ``L"-like correlators}\label{sum}
 \end{figure}

  \begin{figure}[p]
  \begin{center}
   \includegraphics[height=8cm]{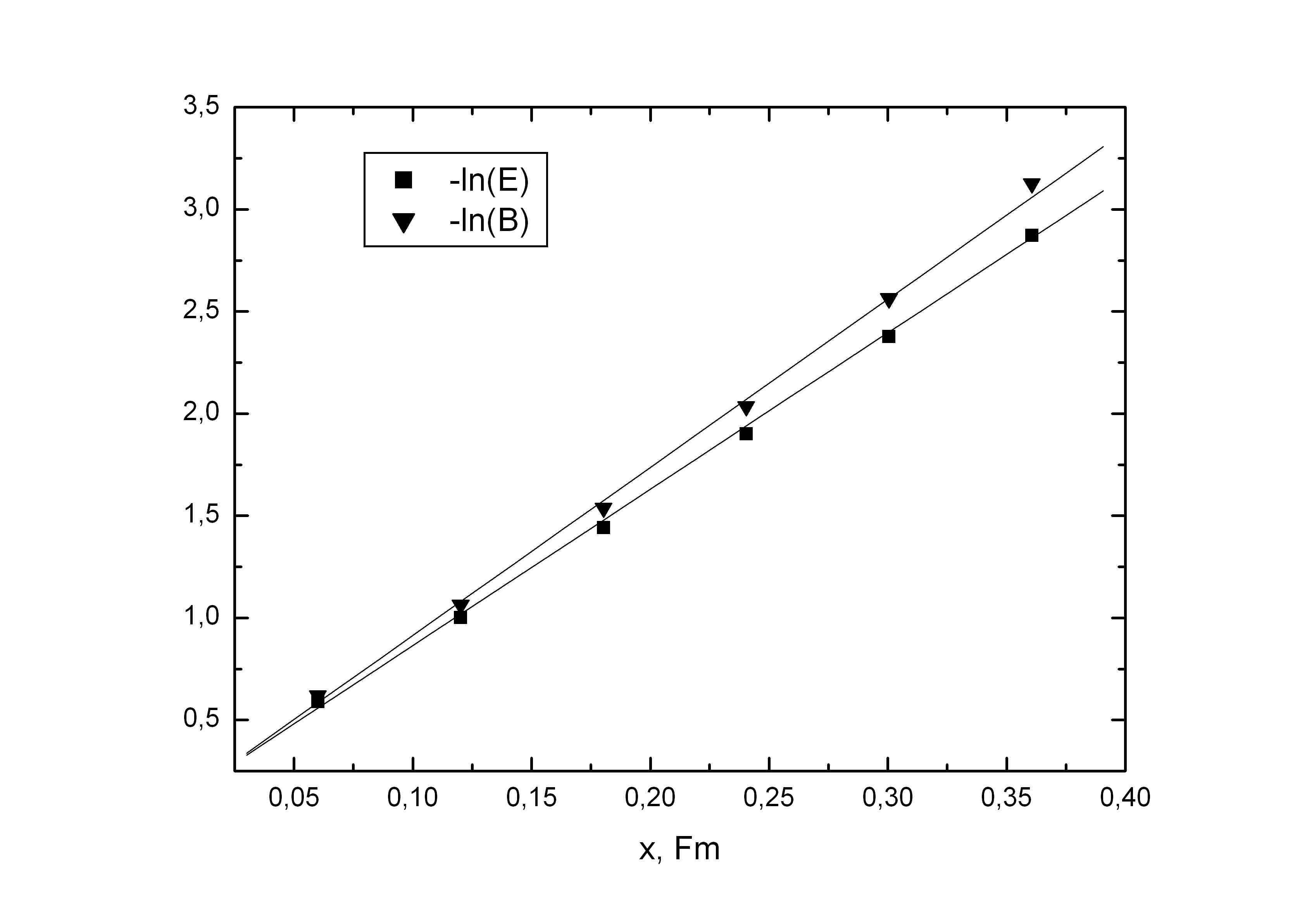}
  \end{center}
  \caption{Correlators of electric and magnetic fields (14)}\label{E&B}
 \end{figure}

\end{document}